\documentclass[mathleft,fleqn,]{an}

\usepackage{amsmath}
\usepackage{graphicx}
\usepackage[varg]{txfonts}
\begin{document}

\Pagespan{1}{}
\Yearpublication{2014}%
\Yearsubmission{2014}%
\Month{0}%
\Volume{999}%
\Issue{0}%
\DOI{asna.201400000}%

\title{Angular momentum, accretion and radial flows in chemodynamical models of spiral galaxies}
\author{G. Pezzulli\inst{1,2}\fnmsep\thanks{Corresponding author:
        {gabriele.pezzulli@phys.ethz.ch}} \and  F. Fraternali\inst{2,3}}
\titlerunning{Angular momentum and abundance gradients}
\authorrunning{G. Pezzulli \& F. Fraternali}
\institute{
ETH Zurich, Institute for Astronomy, Wolfgang-Pauli-Strasse 27, 8093 Zurich, Switzerland
\and
University of Bologna, Department of Physics and Astronomy, Viale Berti Pichat 6/2, 40127 Bologna, Italy 
\and 
University of Groningen, Kapteyn Astronomical Institute, Postbus 800, 9700 AV, Groningen, The Netherlands}

\received{XXXX}
\accepted{XXXX}
\publonline{XXXX}

\keywords{galaxies: abundances -- galaxies: ISM -- galaxies: evolution -- Galaxy: abundances -- intergalactic medium}

\abstract{Gas accretion and radial flows are key ingredients of the chemical evolution of spiral galaxies. They are also tightly linked to each other (accretion drives radial flows, due to angular momentum conservation) and should therefore be modelled simultaneously. We summarise an algorithm that can be used to consistently compute accretion profiles, radial flows and abundance gradients under quite general conditions and we describe illustrative applications to the Milky Way. We find that gas-phase abundance gradients strongly depend on the angular momentum of the accreting material and, in the outer regions, they are significantly affected by the choice of boundary conditions.}

\maketitle

\section{Introduction}
Reconstructing the history of the Milky Way, or of any other spiral galaxy, requires the knowledge of its detailed accretion history, as a function of both time and space. 

Despite the mass assembly in our Universe is relatively well understood on large scales, the details of how primordial gas finds its way from the cosmic web down to the small scales of star forming discs depend on multiple and complex physical processes and the predictions of cosmological hydrodynamical simulations are still not unique in this respect (e.g. \cite{Nelson+13}).

Nonetheless, the observed present-day mass distribution of galaxy discs can be profitably used to put constraints on the spatially resolved accretion history. Following Pitts \& Tayler (1989), one can define the \emph{effective accretion rate surface density} $\dot{\Sigma}_\textrm{eff}$ as the time derivative of the total surface density (in gas and stars) of a disc:
\begin{equation}
\dot{\Sigma}_\textrm{eff} (t, R) := \frac{\partial}{\partial t}(\Sigma_\star + \Sigma_\textrm{gas})(t, R) .
\end{equation}
The effective accretion is strongly constrained by observations, since the integral over time of $\dot{\Sigma}_\textrm{eff}(t, R)$ has to be equal to the total observed surface density at radius $R$. However, to derive the actual accretion rate surface density $\dot{\Sigma}_\textrm{acc}$, $\dot{\Sigma}_\textrm{eff}$ has to be decomposed into its two contributions: direct accretion from the intergalactic medium (IGM) and a radial gas flow within the disc (e.g. \cite{SB09}). The two components are physically linked to each other by angular momentum conservation (\cite{MV81}) and therefore the inference of the accretion profile from the observed mass distribution of the disc depends on the angular momentum distribution of the accreting material (\cite{PT89}; \cite{BS12}).

Since the shape of the accretion profile and the presence of radial flows have a drastic impact on the predicted chemical abundance profiles (e.g. \cite{Tosi88}; \cite{GK92}; \cite{PC00}), it will be of great importance, in future chemodynamical models of the Milky Way, to include a recipe allowing both accretion and radial flows to be computed consistently under general conditions.

In the following, we summarize the general solution to decompose the effective accretion into direct accretion and radial flows, as a function of the angular momentum of the accreting gas (Sec. \ref{sec::theory}), then we give some examples for different angular momentum distributions and two models for the mass assembly of the disc (Sec. \ref{sec::MilkyWay}) and we finally discuss the impact of boundary conditions on the shape of abundance profiles, in particular at the disc periphery (Sec. \ref{sec::BoundaryConditions}). Sec. \ref{sec::summary} sums up.

\section{Accretion, radial flows and chemical evolution}\label{sec::theory}
Following Pezzulli \& Fraternali (2016), the accretion profile is related to the one of effective accretion by:
\begin{equation}
\dot{\Sigma}_\textrm{acc} (t, R) = \frac{1}{R^2 \alpha(t, R) h(t, R)} \int_0^R \hat{R} h(t, \hat{R}) \dot{\Sigma}_\textrm{eff} (t, \hat{R}) d\hat{R}
\end{equation}
with:
\begin{equation}
h(t, R) \equiv \exp \left( \int_{R_0}^R \frac{d\hat{R}}{\hat{R} \alpha(t, \hat{R})}\right)
\end{equation}
where $R_0$ is arbitrary, while:
\begin{equation}\label{defalpha}
\alpha := \frac{V_\textrm{disc} - V_\textrm{acc}}{\partial (RV_\textrm{disc})/ \partial R}
\end{equation}
is a dimensionless measure of the local angular momentum mismatch between the accreting material and the disc, with $V_\textrm{acc}$ and $V_\textrm{disc}$ denoting their respective rotational velocities. Note that $\alpha$, $V_\textrm{acc}$ and $V_\textrm{disc}$ are, in general, functions of $t$ and $R$.

The radial velocity of the gas follows from angular momentum conservation (\cite{MV81}; \cite{LF85}):
\begin{equation}
u^R  = - \alpha R \frac{\dot{\Sigma}_\textrm{acc}}{\Sigma_\textrm{gas}}
\end{equation}
Note that, even in a simple case in which $\alpha$ is constant, the radial velocity will have, in general, a complex dependence on both space and time.

Finally, chemical evolution can be computed, in first approximation, integrating the equation:
\begin{equation}\label{metaleq}
\frac{DX}{Dt} = \frac{\dot{\Sigma}_\star}{\Sigma_\textrm{gas}} - X\frac{\dot{\Sigma}_\textrm{acc}}{\Sigma_\textrm{gas}}
\end{equation}
(where $X$ is the abundance by mass of the considered chemical element, normalized to its integrated yield) along the characteristics lines $(t, R_c(t))$, defined by the equation $dR_c/dt = u^R$. Note that \eqref{metaleq} is linear and it therefore admits an explicit solution:
\begin{equation}\label{metalsol}
X(t, R) = e^{-\sigma(t, R)} \int_0^t e^{\sigma(\hat{t}, R)} \frac{\dot{\Sigma}_\star}{\Sigma_\textrm{g}}(\hat{t}, R) d\hat{t}
\end{equation}
with:
\begin{equation}
\sigma(t, R) :=  \int_0^t \frac{\dot{\Sigma}_\textrm{acc}}{\Sigma_\textrm{g}}(\hat{t}, R) d\hat{t}
\end{equation}

Note that this formalism, based on instantaneous recycling and constant integrated yields, can capture only the general metallicity evolution of the ISM. Individual abundance profiles will slightly differ from one element to another, because of differences in the metallicity dependence of the yields and because of time-delay effects, which are important for elements mainly produced by long-lived stars (e.g. \cite{Romano+10}). Furthermore, the distribution of stellar abundances will be additionally affected by radial migration (e.g. \cite{SB09}). However, gas phase abundances of $\alpha$ elements should be fairly described by \eqref{metalsol} and, as we exemplify in the next Section, they are a valuable tracer of the angular momentum assembly of galaxy discs.

\section{Application to the Milky Way}\label{sec::MilkyWay}
\begin{figure*}
\centering
\includegraphics[width=17.1cm]{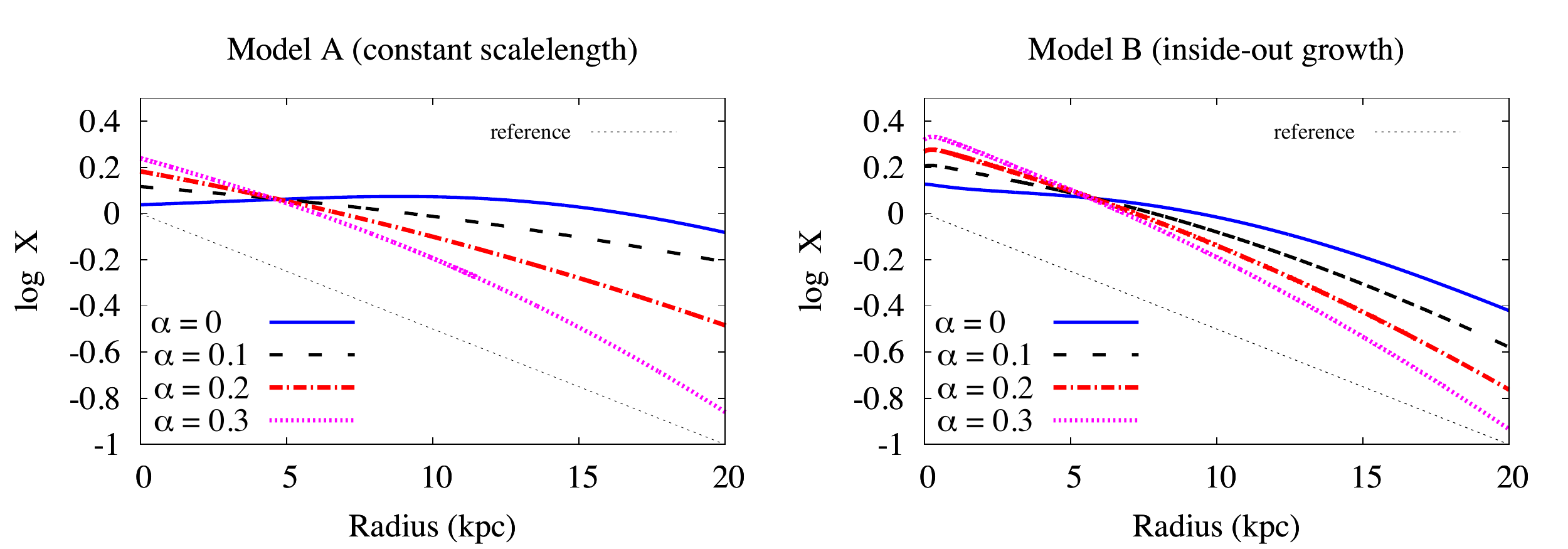}
\caption{Gas-phase abundance profiles of $\alpha$-elements (in yield units), for different values of the angular momentum mismatch parameter $\alpha$ and two structural evolution models of the Milky Way. Model A (\emph{left panel}) assumes a constant stellar disc scalelength. Model B (\emph{right panel}) is an inside-out growing model. The `reference' straight line, put to guide the eye, has an arbitrary normalization and a slope of $-0.05 \; \textrm{dex}/\textrm{kpc}$, equal to the median abundance gradient of $\alpha$-elements in Milky Way Cepheid stars, mostly at galactocentric distances between 5 and 15 kpc (\cite{Genovali+15}).}\label{fig::Fig1}
\end{figure*}
From Sec. \ref{sec::theory}, the key ingredients to compute abundance profiles are the structural evolution of the disc (parametrized by $\dot{\Sigma}_\textrm{eff}$) and the angular momentum of the accreting material (quantified by $\alpha$). To illustrate their effect on chemical evolution, we show in Fig. \ref{fig::Fig1} the predictions of models with varying $\alpha$ (which for simplicity we assume here to be independent of time and radius \footnote{This is the case, for instance, if both the disc and the accreting material have a flat, time-independent rotation curve, see equation \eqref{defalpha}.}) and for two structural evolution models of the Milky Way. In both cases, we assume an exponential stellar disc, formed according to the Kennicutt-Schmidt law (\cite{K98}) and an exponential star formation history:
\begin{equation}
\dot{M}_\star (t) = \frac{M_{\star, \infty}}{t_\star} \exp \left( -\frac{t}{t_\star} \right)
\end{equation}
where the parameters $M_{\star, \infty} = 6.5 \times 10^{10} \; \textrm{M}_\odot$, $t_\star = 12.5 \; \textrm{Gyr}$ are chosen to yield, after 12 Gyr, a disc stellar mass of $4 \times 10^{10} \; \textrm{M}_\odot$ (\cite{DB98}) and a net star formation rate of $\dot{M}_\star = 2 \; \textrm{M}_\odot \; \textrm{yr}^{-1}$ (\cite{Misiriotis+06}; we assume that a fraction $\mathcal{R} = 0.3$ of the stellar mass is returned to the ISM). In model A, we assume the scalelength to be constant and equal to 2.5 kpc (\cite{Juric+08}). In model B, the scalelength is allowed to grow with time (\emph{inside-out growth}). As shown by Pezzulli et al. (2015), this implies a present-day radial profile for the star formation rate surface density of the form:
\begin{equation}\label{SFRD}
\dot{\Sigma}_\star (R) = \frac{M_\star}{2 \pi R_\star^2} \left( \nu_M + \nu_R \left( \frac{R}{R_\star} - 2 \right) \right) \exp \left( - \frac{R}{R_\star} \right)
\end{equation}
where:
\begin{equation}
\nu_M := \frac{\dot{M}_\star}{M_\star}
\end{equation}
and
\begin{equation}
\nu_R := \frac{\dot{R}_\star}{R_\star}
\end{equation}
are the \emph{specific mass growth rate} and the \emph{specific radial growth rate} of the stellar disc, respectively. Here we adopt $\nu_M = 5 \times 10^{-2} \; \textrm{Gyr}^{-1}$, $\nu_R = 2.15 \times 10^{-2} \; \textrm{Gyr}^{-1}$, $R_\star = 2.43$ kpc, which are compatible with the most recent observed distribution of supernova remnants in the Milky Way (\cite{Green15a},  2015b).  Furthermore, we assume the ratio between $\nu_M$ and $\nu_R$ to be constant with time.

In both Model A and Model B, the steepness of the abundance gradient strongly increases with increasing $\alpha$. For any fixed value of $\alpha$, Model B shows a steeper gradient than Model A, in agreement with expectations from classical studies of inside-out growth (e.g. \cite{Molla+97}; \cite{BP99}; \cite{Chiappini+01}). A moderate, but non-vanishing, angular momentum mismatch is required, in both cases, to produce a gradient similar to the one measured from Cepheid stars (which are a good tracer of the ISM composition) in the Milky Way (\cite{Genovali+15}).

\section{Dependence on boundary conditions}\label{sec::BoundaryConditions}
\begin{figure*}
\centering
\includegraphics[width=17.1cm]{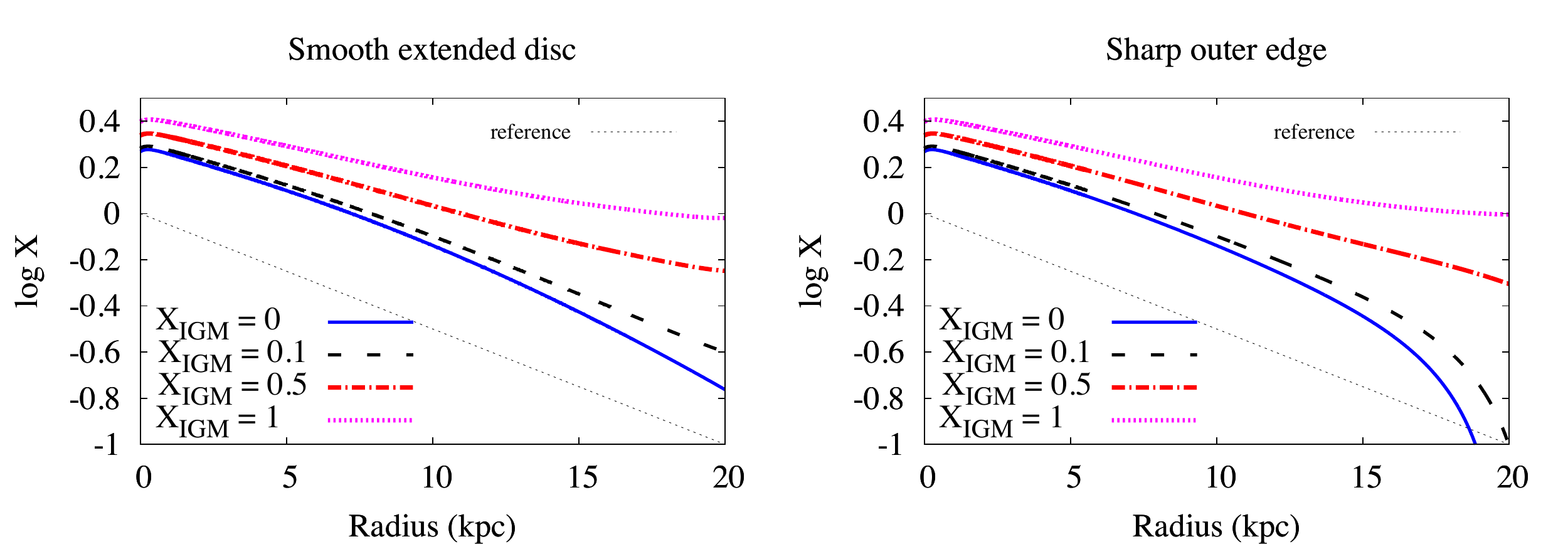}
\caption{Dependence of abundance profiles on boundary conditions, for an inside-out model (model B) with $\alpha = 0.2$ (see Sec. \ref{sec::MilkyWay} and Fig. \ref{fig::Fig1}). \emph{Left panel:} the case of a smooth disc, extending out to $R = 30$ kpc or more. \emph{Right panel:} predictions if the disc has an outer edge at $R = 20$ kpc, from which radial accretion directly occurs from the IGM. Different lines are for different metallicities of the IGM; the reference line is as in Fig. \ref{fig::Fig1}.}\label{fig::Fig2}
\end{figure*}

The method of characteristics (Sec. \ref{sec::theory}) is very effective to trace the impact of different boundary conditions on abundance profiles (see also \cite{PF16}) and in particular on their shape in the outer regions, where models often disagree with each other and with observations.

The results presented in the previous Section (Fig. \ref{fig::Fig1}) are computed assuming primordial accretion onto a disc smoothly extending out to at least $R = 30$ kpc. In Fig. \ref{fig::Fig2}, we illustrate the effects of a non primordial composition of the accreting material and of a sharp disc edge from which direct radial accretion from the IGM is allowed to occur. We consider an inside-out model with $\alpha = 0.2$. The normalized IGM metallicity $X_\textrm{IGM}$ is assumed to linearly increase with time, up to a maximum value between 0 (primordial composition) and 1 (pre-enrichment equal to the integrated yield). Left panels are for a smooth disc (similar to Fig. \ref{fig::Fig1}), while right panels are for direct radial accretion occurring from an outer edge at $R = 20$ kpc.

A non vanishing metallicity of the IGM has the obvious effect of increasing the general level of chemical enrichment in the ISM, but it also has some impact on the shape of metallicity distributions. In the smooth disc models, abundance profiles always gently decline with radius, with the tendency to become shallower with increasing $X_\textrm{IGM}$. In contrast, models with direct radial accretion show a wider variety of behaviours, ranging from a very strong steepening (for primordial IGM composition) to an outer flattening (for high enough IGM metallicity).

Note that the enrichment of the accreting gas can be partially due to mixing of the IGM with metals produced in the disc itself and then ejected in the halo by AGN or supernova feedback.

\section{Summary}\label{sec::summary}
The spatially resolved accretion history of the Milky Way is a fundamental ingredient for any chemical evolution model of our Galaxy. However, its reconstruction from observations strongly depends on the detailed angular momentum distribution of the accreting material, which is still largely unknown and likely to be quite complex.

We provided a summary of the analytic methods to derive both the accretion profile and the radial velocity of the gas, consistently with angular momentum conservation and under fairly general conditions, which will be useful in future detailed chemodynamical models of the Milky Way.

We discussed how these profiles can be used to infer the gas-phase distribution of $\alpha$-elements. We presented illustrative abundance profiles derived for the Milky Way, in both an inside-out and a constant scale-length framework, and we have found them to be very sensitive to the angular momentum distribution of the accreting material.

We finally addressed the dependence of the results on boundary conditions such as the metallicity of the accreting material and the presence of an outer edge of the disc. The latter ingredient in particular can have a very strong impact on the steepness of abundance gradients at large galactocentric distances.

\acknowledgements
We acknowledge Dave Green for kindly making and sharing with us some supplementary analysis on the distribution of supernova remnants in the Milky Way. We also acknowledge financial support from PRIN MIUR 2010-2011, project `The Chemical and Dynamical Evolution of the Milky Way and Local Group Galaxies', prot. 2010LY5N2T.

\end{document}